\documentclass[a4paper,twoside]{article}

\usepackage{subfig}
\usepackage{epsfig}
\usepackage{calc}
\usepackage{float}
\usepackage{amssymb}
\usepackage{amstext}
\usepackage{amsmath}
\usepackage{amsthm,amsfonts}
\usepackage{multicol}
\usepackage{pslatex}
\usepackage{apalike}
\usepackage{hyperref}

\def\BibTeX{{\rm B\kern-.05em{\sc i\kern-.025em b}\kern-.08em
    T\kern-.1667em\lower.7ex\hbox{E}\kern-.125emX}}

\usepackage{SCITEPRESS} 

\begin{document}

\title{Pervasive Hand Gesture Recognition for Smartphones using Non-audible Sound and Deep Learning}

\author{\authorname{
Ahmed Ibrahim \orcidAuthor{0000-0003-0898-1096},
Ayman El-Refai  \orcidAuthor{0000-0003-1996-3774},
Sara Ahmed \orcidAuthor{0000-0003-4367-6628},  
Mariam Aboul-Ela  \orcidAuthor{0000-0002-2998-2197},
Hesham M. Eraqi \orcidAuthor{0000-0001-9430-7553},
Mohamed Moustafa \orcidAuthor{0000-0002-0017-3724}
}

\affiliation{Department of Computer Science and Engineering, The American University in Cairo, New Cairo, Egypt}

\email{\{ahmeddibrahim, ayrefaie, sabouzeid0, mariam-a, heraqi,  m.moustafa\}@aucegypt.edu}
}


\abstract{Due to the mass advancement in ubiquitous technologies nowadays, new pervasive methods have come into the practice to provide new innovative features and stimulate the research on new human-computer interactions. This paper presents a hand gesture recognition method that utilizes the smartphone's built-in speakers and microphones. The proposed system emits an ultrasonic sonar-based signal (inaudible sound) from the smartphone's stereo speakers, which is then received by the smartphone's microphone and processed via a Convolutional Neural Network (CNN) for Hand Gesture Recognition. Data augmentation techniques are proposed to improve the detection accuracy and three dual-channel input fusion methods are compared. The first method merges the dual-channel audio as a single input spectrogram image. The second method adopts early fusion by concatenating the dual-channel spectrograms. The third method adopts late fusion by having two convectional input branches processing each of the dual-channel spectrograms and then the outputs are merged by the last layers. Our experimental results demonstrate a promising detection accuracy for the six gestures presented in our publicly available dataset with an accuracy of 93.58\% as a baseline.}

\onecolumn \maketitle \normalsize \setcounter{footnote}{0} \vfill

\section{\uppercase{INTRODUCTION}}
\label{sec:introduction}
Scientists are always interested in revolutionizing human-machine interaction, also known as HMI. As a result, modern technologies have been introduced that are often placed in advanced technologies, such as the LiDAR (Light Detection and Ranging) sensor in the iPhone 12. These advanced technologies are often expensive and require the latest technology, often making them inaccessible to the public. 

Our proposed solution is to have an acoustic active sensing system to classify the various hand gestures a user can create. The meaning of active sensing is that the sound signal is emitted and received by the same smartphone \cite{survey}. In such a case, the smartphone’s built-in speakers transmit a sound signal at a specific frequency and record the changes in the signal from the hand movement around the smartphone based on the Doppler Effect. Generally, for gesture recognition applications, systems adopt sound signals with a frequency greater than 16 kHz, as signals higher than 16 kHz are above the average human’s audibility \cite{ultrasonic}. Such signals are referred to as ultrasonic signals. Ultrasonic signal has received a lot of attention recently in many applications due to the following reasons. Firstly, it has been thoroughly studied and applied on many applications, as it has good ranging accuracy and low-cost deployment. Secondly, ultrasound is non-audible to human hearing; therefore, it can be used without interfering with a person’s normal life. Thirdly, such sound signals are not affected by lighting conditions; therefore, the system can work during the day or night without interruption, unlike other sensors, for example, cameras. Therefore, we used an ultrasonic wave to detect hand gestures.

Additionally, our system records the signal while simultaneously acting as an active sonar system. By recording the signals, we apply STFT (Short-time Fourier transform) on the waveform to convert the time domain into a frequency domain; hence we can visualize the hand movement gesture as a Doppler Effect on the spectrogram. Afterward, we utilize deep learning techniques to classify hand gestures. We use different types of CNN models with variations in the input. However, all CNN models take the input as a spectrogram. The first model takes the dual-channel audio as 1 channel image of the spectrogram (Basic CNN). The second takes two spectrogram inputs, one as the top microphone and the other as the bottom microphone, and then the data is concatenated and inputted into the model (Early Fusion). The third takes two spectrograms inputs of the top and bottom channels as two separate inputs, and then each input is trained on a different branch then the branches are merged at the end of the model (Late Fusion).

This paper is organized as follows. Section 2 introduces the previous studies on gesture recognition using non-audible frequencies and tracking finger positions. Section 3 discusses the signal generation. Section 4 explains the data collection and augmentation procedures. Also, it shows the dataset components. Section 5 describes the proposed Deep Learning models. Section 6 shows the results and accuracies achieved. Section 6 shows the results of our experiment. Finally, section 7 describes the conclusion and highlights the future work.

\section{\uppercase{RELATED WORKS}}
Multiple studies have been conducted discussing using a smartphone's microphone and speaker to act like a sonar system. \cite{hindawi} propose a system that uses two smartphones. One smartphone generates an ultrasonic wave of 20 kHz, and the other uses the microphone to pick up the signals transmitted by the first phone. After recording the emitted sound during performing hand gestures, and based on the Doppler effect, these recorded signals varied depending on time according to the hand movement or its position. Such a recorded sound signal is then reflected as an image using the short-time Fourier transforms and applied to the convolutional neural network model (CNN) to classify certain hand gestures.
Kim et al.reached an accuracy of 87.75 percent.
As previously mentioned, the reflected paper demonstrates hand gesture recognition by using two smartphones, one as a microphone and the other as a speaker. On the other hand, our research will concentrate on using a single smartphone to perform the gesture recognition process. On the other hand, another technique named UltraGesture \cite{ultragesture}, avoids the Doppler Effect methodology in tracking slight finger motions and hand gestures. Instead, it uses a Channel Impulse Response (CIR) as a gesture recognition measurement with low pass and down conversion filters. In addition, they perform a differential operation to obtain the differential of the CIR (dCIR). Then the information is used as the input data. Lastly, as paper \cite{hindawi}, the authors transform the data into an image trained in a CNN model. However, in the experimental phase, the authors added additional speakers kit on the smartphone, which boomed the accuracy results to an average of 97\% for 12 hand gestures.

Another interesting study is AudioGest \cite{audiogest} which uses one built-in microphone and speaker in a smartphone. However, it does not rely on any deep learning model. Instead, it relies on filters and signal processing. It can recognize various hand gestures, including directions of the hand movement, by using an audio spectrogram to approximate the hand in-airtime using direct time interval measurement. Also, it can calculate the average waving range of the hand by using range ratio and the hand movement speed by using speed-ratio, \cite{audiogest}. According to the authors, AudioGest is nearly unaffected by human noise through its denoising operation and signal drifting issues, which were tested in various scenarios such as in a bus, office, and café. For the results, AudioGest achieves slightly better results in gesture recognition compared to the former \cite{hindawi} which includes recognition in noisy environments.

\begin{figure*}[!ht]
\centering
\vspace{-0.2cm}
\includegraphics[width=8cm,height=8cm,keepaspectratio]{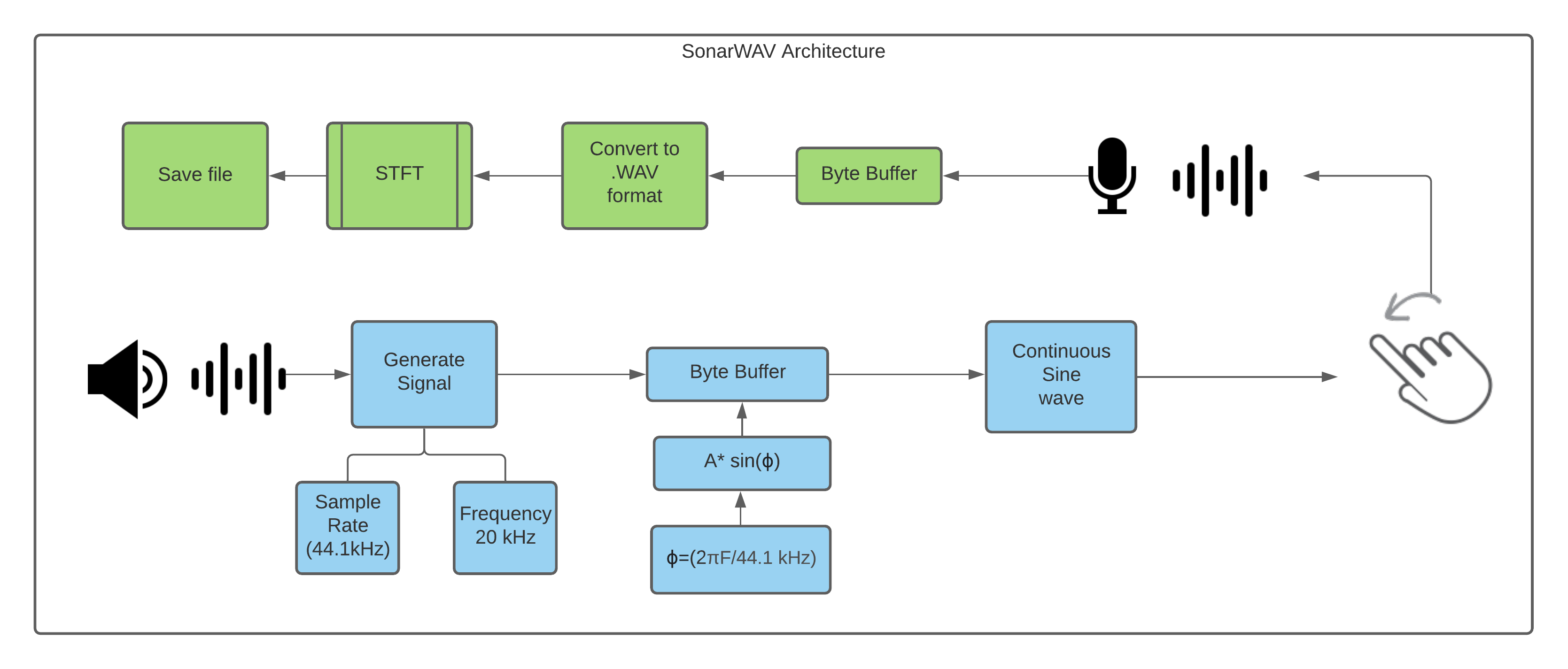}
\caption{Data-flow Pipeline}
\label{sonar}
\vspace{-0.1cm}
\end{figure*}

Sonar sensing can also be used as a finger tracking system. FingerIO \cite{fingerio} presents a finger tracking system to allow users to interact with their smartphone or a smartwatch through their fingers, even if it is occluded or on a separate plane from the interaction surface. The device’s speakers emit inaudible sound waves 18-20 kHz, then the signals are reflected from the finger and are recorded through the device’s microphones. For the sake of relevancy, the research will focus on smartphones only. FingerIO’s operation is based on two stages, transmitting the sound from the mobile device’s speakers, then measuring the distance from the moving finger by the device’s microphone. For the speaker’s part, it uses Orthogonal frequency division multiplexing (OFDM) to accurately identify the echo’s beginning to improve the finger tracking accuracy. For the results, without occlusions, FingerIO attains an average accuracy of 8mm in 2D tracking. The interactive surface works effectively (error within 1 cm) within a range of 0.5 m2. Exceeding the range increases the margin of error to 3 cm. However, in occlusions, the 2D tracking average accuracy rises to 1 cm. Besides, FingerIO requires an initial gesture before attempting any gesture to avoid false positives, which worked 95\% of the time.
Nevertheless, FingerIO \cite{fingerio} suffers from several limitations such as: Tracking multiple concurrent motions, a High margin of error when interfering within 50cm of the device, and Tracking 3-D finger motion. 

\section{\uppercase{SIGNAL GENERATION}}

\subsection{Background}
This section provides a concise overview of the processing system for recognizing gestures activities using a smartphone's ultrasonic signal. In the acoustic sensing system, waveform generation plays an important role because it specifies the signal characteristics. To get the benefit of the design of a good waveform, we perform denoising and feature extraction. Many types of sound signals can be used according to the characteristics of sound waveforms \cite{signals}. We consider three types of sound signals which are OFDM, Chirp, and CW. The main difference between the first two is that the OFDM uses frequencies emitted multiple times during a specific time frame, making it harder to process, and causes the signal complexity to be quite high. However, the Chirp signal uses a single frequency from a range of frequencies and keeps increasing linearly. Also, the frequencies in this range should be below 20 kHz; however, using a frequency below 20 kHz is not adequate since it will be audible. \cite{Chirp}.

\begin{figure}[!ht]
  \centering
  \subfloat[Before filtering]{\includegraphics[width=0.25\textwidth]{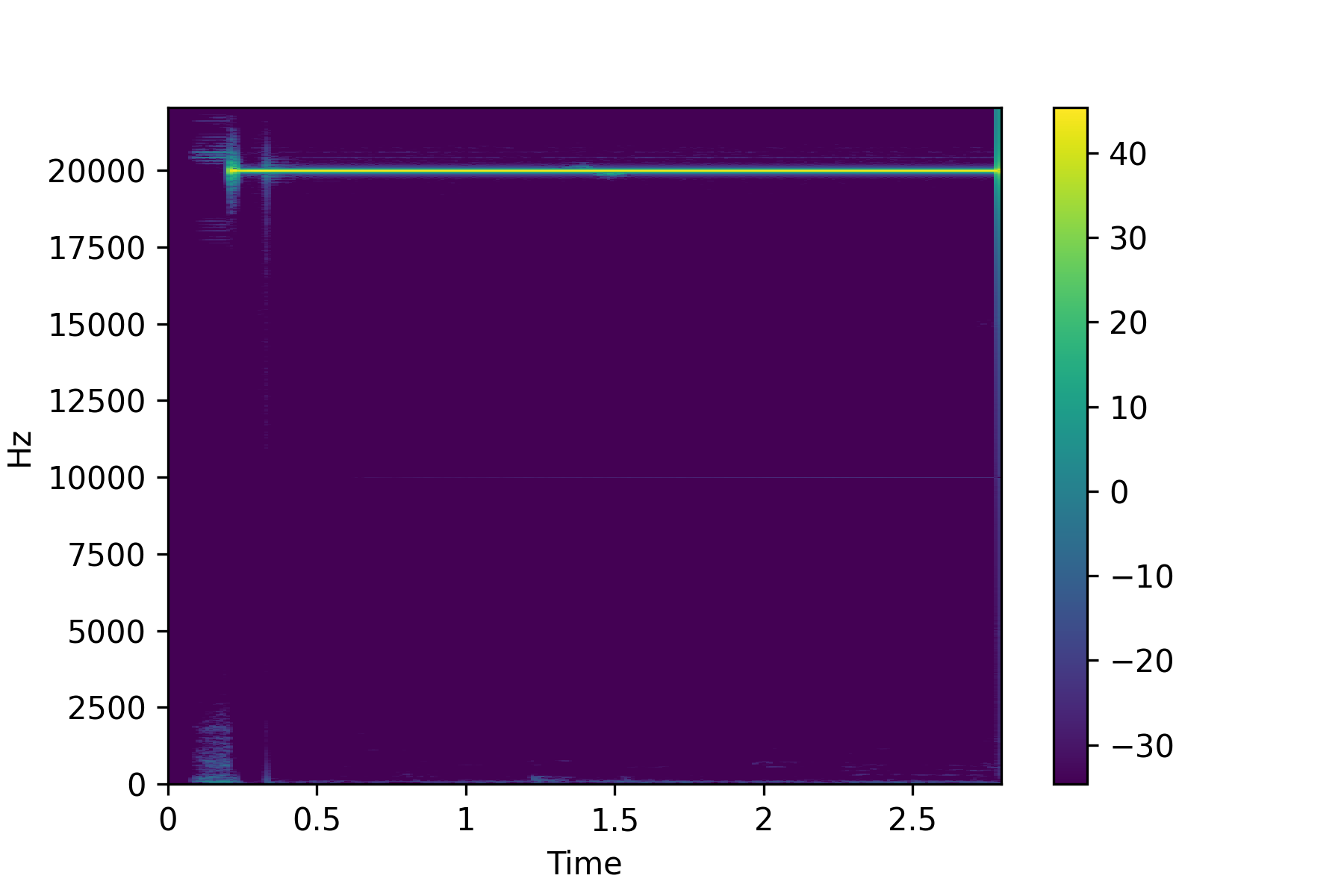}\label{fig:f1}}
  \subfloat[After filtering]{\includegraphics[width=0.25\textwidth]{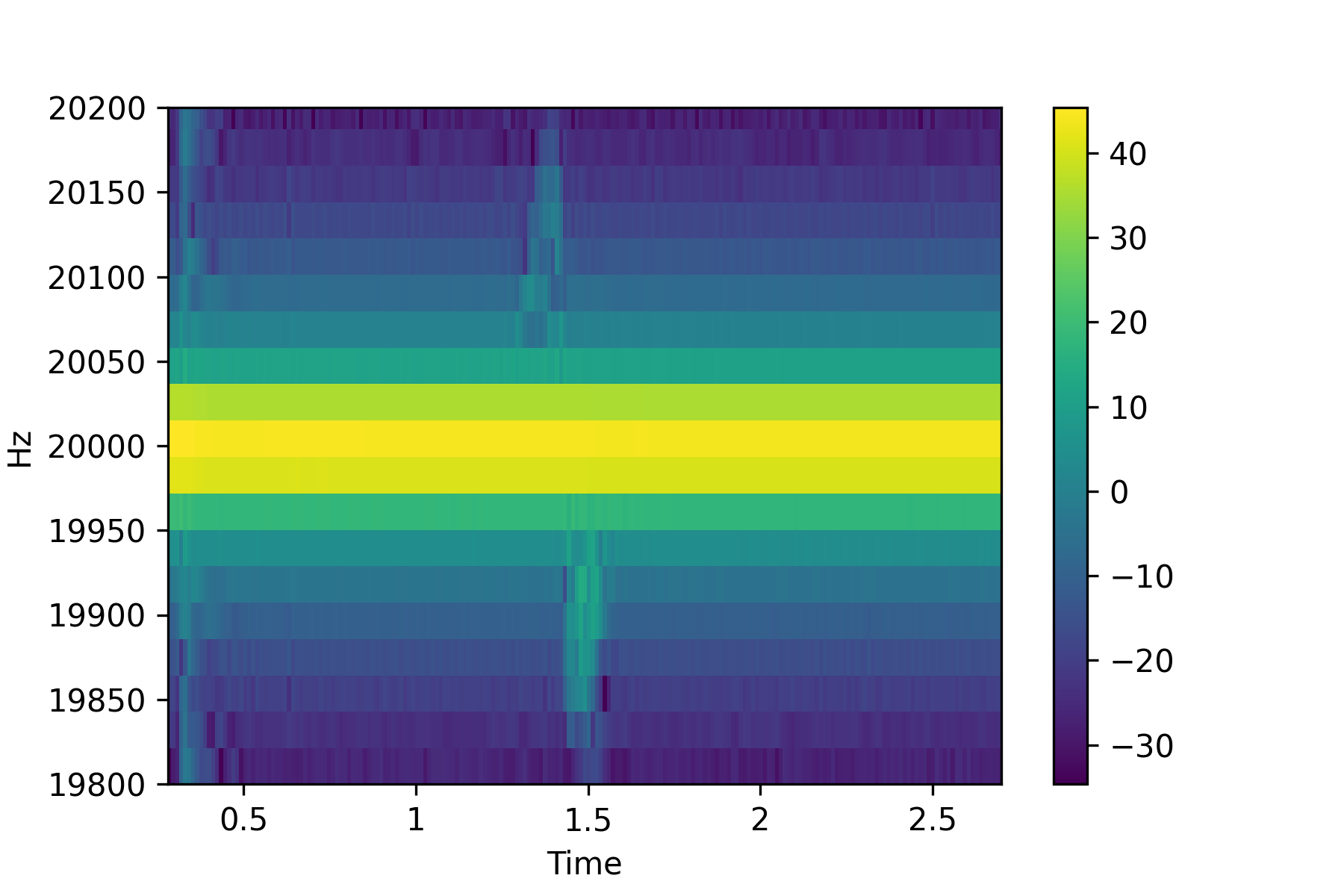}\label{fig:f2}}
  \caption{Filtering the spectrogram into a specific frequency band}
\end{figure}

\subsection{Continuous Wave}
The last type is continuous wave (CW) which is the signal that we decide to generate. It is much more vulnerable to background noises than OFDM and does not work well with long distances, yet these drawbacks do not affect our system (More details in sections \textbf{3.3} and \textbf{3.4}). This wave uses a single frequency emitted over a period of time. This makes it great as it does not consume a lot of power and makes the processing easier. When the transmission energy in each frequency is reduced, the Signal-to-Noise Ratio (SNR) is also reduced. We have also found that the CW signal is often consistent with little to no signal loss and less noise due to the narrower bandwidth compared to the OFDM. These are all reasons that lead us to use CW as generated wave signal.

\subsection{Signal Generation Procedure}
As mentioned, our system uses one smartphone without extra hardware. We utilize both speakers and microphones at the top and bottom (two emitters and two receivers). 
We generated a CW at 20 kHz frequency and a sample rate of 44.1 kHz using the two speakers built-in the smartphone (top and bottom) using equation \eqref{eq1} where F is the frequency (20 kHz) and the input changes based on the current sample in the audio. Afterward, the output is saved in a buffer and adjusted to produce 16-bit PCM.

\begin{equation}\label{eq1}
    f( x) \ =\sin{( F\ *\ 2\pi \ *\ x)}
\end{equation}

\begin{equation}\label{eq2}
    f\ =\ \frac{f( v\ +\ v_{o})}{( v-v_{s})}
\end{equation}

After the signal is generated, a hand gesture is performed in front of the smartphone. Then a shift is recorded according to the movement caused by the Doppler Effect by equation \eqref{eq2}, where $v_{o}$ represents the velocity of the observed object (hand) and $v_{s}$ represents the velocity of the source (smartphone). Then it is recorded by the two built-in microphones in the smartphone. Figure \ref{sonar} shows the entire process.

After acquiring the recording audio in .wav file format, we applied Short-time Fourier transform (STFT) on the data to display a spectrogram showing the generated frequencies. The STFT frequency resolution was set as 2048. After acquiring the spectrogram, we filtered it to obtain the essential parts for classifying the gestures. First, we narrowed down the spectrogram duration from 1.3 to 2.7 seconds. Therefore our spectrogram image only includes a section that is 1.4 seconds long. Then, we filtered the frequency band from 19.7 to 20.3 kHz, thus removing the effect of audible noise without adding a band-pass filter, as shown in Figures \ref{fig:f1} and \ref{fig:f2}.

\subsection{Noise Tolerance}
\raggedbottom
Looking into the factors that would allow our system to be able to tolerate as much noise as possible. We were able to discern two different types of noise; high-frequency noise and low-frequency noise. We've defined high-frequency noise as any frequency over 15 kHz, while low-frequency noise is anything below that threshold. We aimed to successfully mitigate the low-frequency noise so that any background noise such as talking or movement away from the microphone would not be able to harm the user's experience. We were able to successfully mitigate that by taking a frequency range of 19 kHz to 23 kHz. That way we were able to mitigate any of the noise that is below 19 kHz. We were also able to increase noise tolerance by using data augmentation with noise injection (More details in section \textbf{4.2.1}).

\section{\uppercase{DATA}}

\subsection{Data Collection}
\begin{figure}[htbp]
\begin{center}
\includegraphics[width=4cm,height=8cm,keepaspectratio]{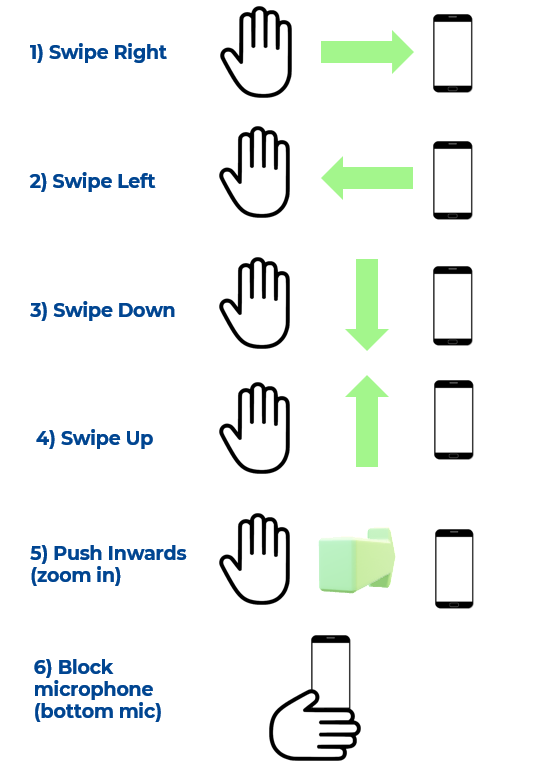}
\end{center}
\caption{The  six  gestures in spectrogram image.  1-  Swipe  Left,  2- Swipe  Down,  3-  Push Inwards,  4- Swipe  Right,  5- Swipe Up,  6- Block Microphone}
\label{gestures}
\end{figure}

We gathered signal reflections from 6 predetermined gestures using a Samsung Galaxy S10e smartphone, each of the gestures generating a different pattern due to the Doppler Effect. This causes the data collection to be a lot easier and ensures the integrity of the data. The 6 supported gestures are 1) Swipe Right, 2) Swipe Left, 3) Swipe Down, 4) Swipe Up, 5) Push Inwards, 6) Block the Microphone as in Figure \ref{gestures}. We recorded the signal reflections as WAV files and had initially converted each WAV file to a single spectrogram. In total, we had  6 types of gestures from 4 separate users, each of which performed 80 gestures amounting to 1920 collected gestures.

However, upon further investigation, we realized that the single spectrogram generated from the WAV file could be further broken down into two spectrograms, one generated from a smartphone's top speaker and microphone and the other from the bottom speaker and microphone of the smartphone. We used the FFmpeg software to help us separate such data \cite{ffmpeg}. This obviously doubled our dataset and allowed us to further experiment with different types of CNN models, as well as provide better accuracy for our model.

\subsection{Data Augmentation}
Data augmentation includes several techniques that improve the size and quality of the training datasets, allowing them to build better Deep Learning models. We worked on two different types of data Augmentation, which are Raw Audio and spectrogram image.

\subsubsection{Raw Audio Augmentation} Regarding augmenting the raw audio, we decided to implement noise injection to test how resistant our model was to external noises. Also, we apply frequency and time masking to the dataset.
\begin{enumerate}
    \item \textbf{Noise Injection:} Concerning the noise injection, We did so by injecting some random value within a range into the time series data as shown below in Figure \ref{inject}. We decided to use such a method as we found that overlaying two sounds had little effect on the spectrogram itself, as most audio found online had a lower frequency than the one we were using.  

\begin{figure}[htbp]
\begin{center}
\includegraphics[width=7cm,height=8cm,keepaspectratio]{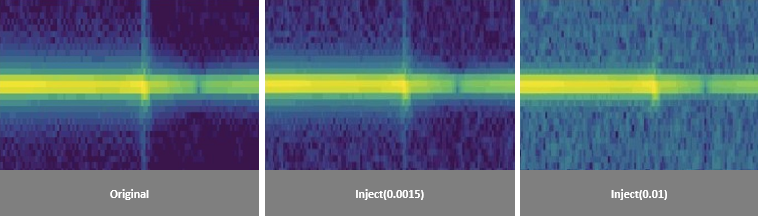}
\end{center}
\caption{Image After Noise Injection}
\label{inject}
\end{figure}

\item \textbf{Frequency and Time Masking:} 
Regarding raw audio manipulation, we also explored the idea of adding frequency and time masking to the raw audio provided. We took the raw audio with frequency masking and added a lower frequency that was much stronger than the one found in our raw audio. Through that, the stronger and lower frequency will effectively mask our original audio's frequency. While this method was great in theory, in practice, it proved to be much more difficult. The issue was that the frequency generated was often much lower, even through continuous adjustments. This meant that it did not affect the original WAV file, and therefore was not included in our augmentation process. The same is true for time masking, but the mask was placed on the time domain, which was not helpful as we attempted to identify the change in the frequency signal.

\end{enumerate}

\subsubsection{Spectrogram Images Augmentation}
Moving to the data augmentation using spectrogram images, the image augmentation algorithms discussed in this section include translation and Gaussian Noise.

\begin{enumerate}
    \item \textbf{Translation:} shifting images left, right, up, or down is useful to avoid personal bias in the data. However, this type is useful if all the images in the dataset are centered, yet the data can occur at any part of the image in our dataset. Therefore, we worried that simple-shifting might destroy the data as a whole. So, we tried to apply a simple shift as in Figure \ref{original_vs_after_shift:f1} compared to the original in Figure \ref{original_vs_after_shift:f2}, we only apply random width shifts of the slandered deviation to avoid cutting the data, yet it was not very effective. Also, we believe that large shifts should be more harmful as they could cut the data.

\begin{figure}[!ht]
  \centering
  \subfloat[Example of Original spectrogram Image]{\includegraphics[width=0.2\textwidth]{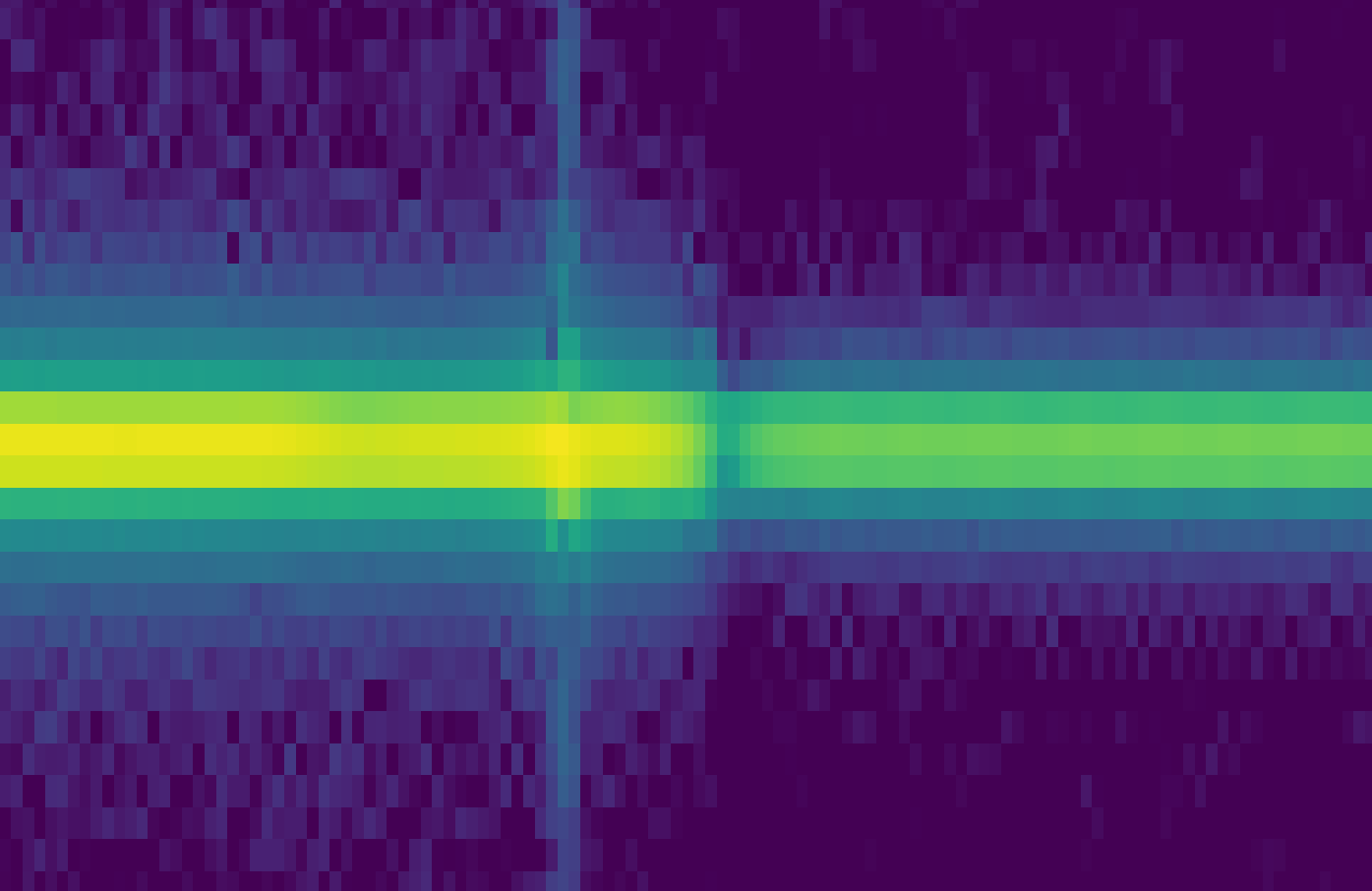}\label{original_vs_after_shift:f1}}
  \hfill
  \subfloat[Image After Random Shift]{\includegraphics[width=0.2\textwidth]{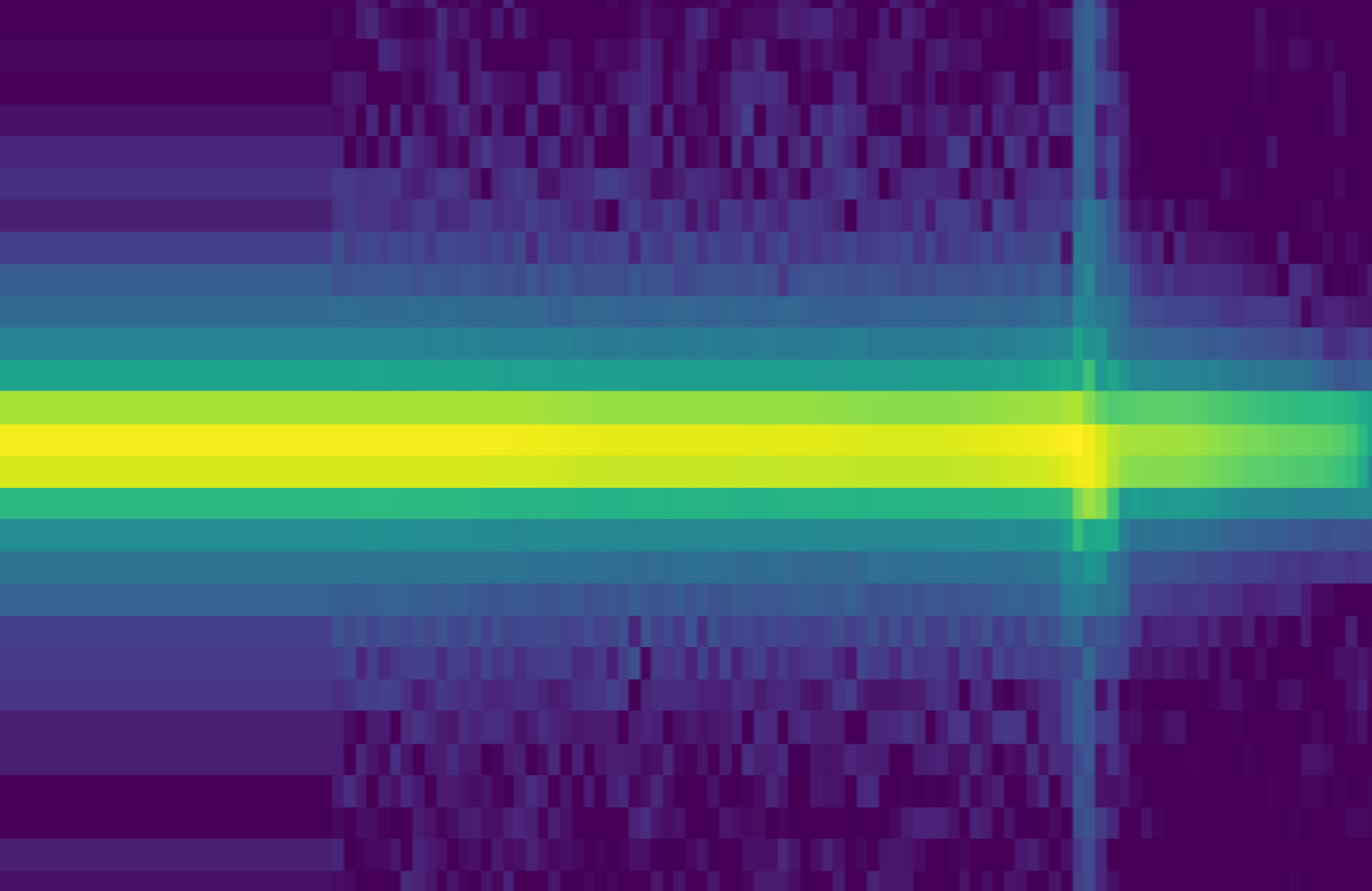}\label{original_vs_after_shift:f2}}
  \caption{Difference between original spectrogram image before and after random shift}
\end{figure}

\begin{figure}[h]
\begin{center}
\includegraphics[width=7cm,height=8cm,keepaspectratio]{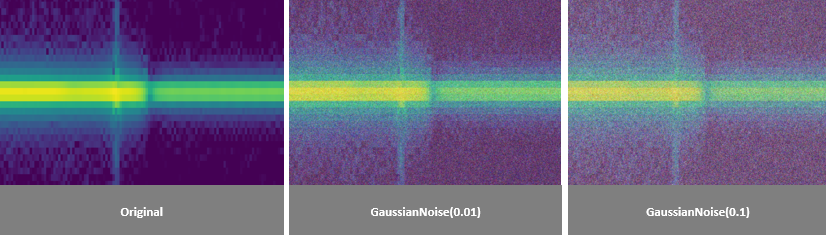}
\end{center}
\caption{Gaussian Noise}
\label{Gauss}
\end{figure}

\begin{figure*}[!ht]
\includegraphics[width=1.0\textwidth]{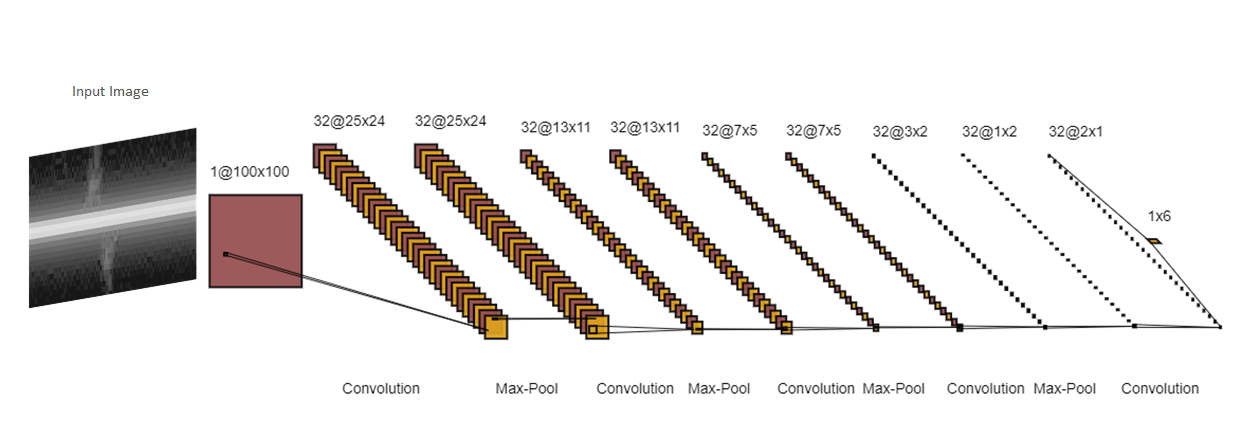}
\caption{Our original CNN model architecture}
\label{cnn_org}
\end{figure*}

\item \textbf{Gaussian Noise}: is statistical noise having a probability distribution function that is equal to that of the normal distribution. It is also known as the Gaussian distribution. According to \cite{gaussian} Gauss, events are prevalent in nature; while observing random events, which is the sum of many independent events, all random variables appear to be Gauss variables. This is the reason why Gaussian distribution is extremely preferred in machine learning.
Gaussian noise can be applied to the dataset through this method. First, we calculate random noise and assign it to the variable. Then it has been added to the dataset. To calculate the Gaussian distribution probability density function using the following formula \eqref{eq3}:

\begin{equation}\label{eq3}
    N( x:M,\sigma ) \ =\ \frac{1}{\sqrt{2\pi \sigma ^{2}}} exp\left( -\frac{1}{2}( x-M)^{2} /\sigma ^{2}\right)
\end{equation}

We used the random noise function from the skimage package to add the Gaussian noise to augment the dataset before training the model. The function requires adding the image, the type of noise that is Gaussian noise and, the variance of random distribution, which is equal to the double of the stander deviation. So, we tried different values of variance to obtain clear images with Gaussian noise. It found that the smaller the variance, the clearer the image with Gaussian noise, as shown in Figure \ref{Gauss}. Hence, we chose the values of (0.01) not to damage the spectrogram's data.
\end{enumerate}

\begin{figure}[b]
\centering
\centerline{\includegraphics[width=8cm,height=8cm,keepaspectratio]{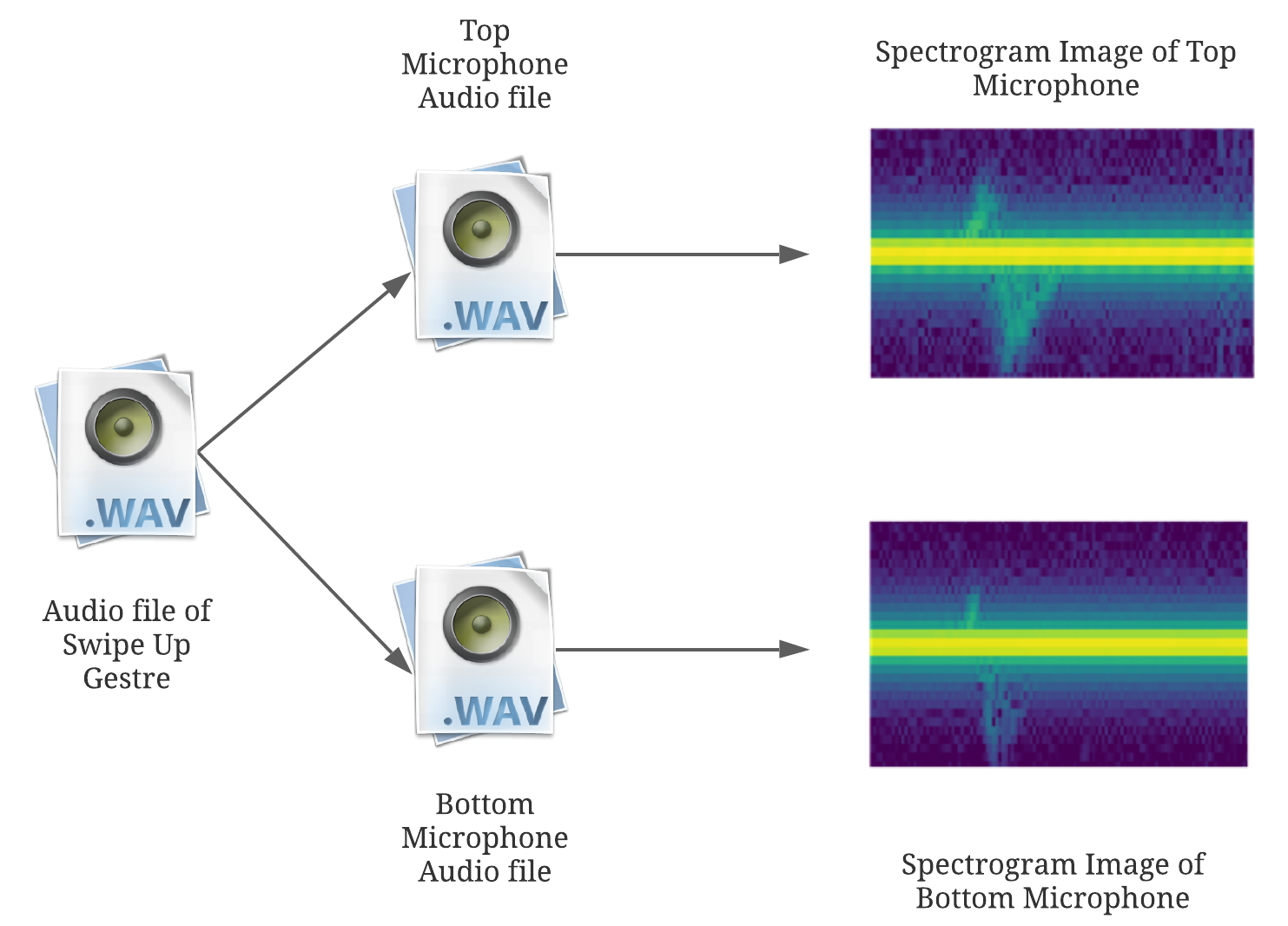}}
\caption{Separating Audio file into two audio files then converting the audio to two spectrogram images}
\label{sep_audio}
\end{figure}

\subsection{Dataset}

The data were collected in noisy and non-noisy environments. The noise included sound coming from the air conditioner, people conversations, and music as well. The data is divided into an audio file (.WAV) and spectrogram images. The spectrogram images were filtered between 19.7 and 20.3 kHz, as this is the range where the sonar and gesture are shown. It also removes the noise coming from lower frequencies without needing to use a bandpass filter. Moreover, the data were recorded as a stereo; thus, it contains the two channels of the top and bottom microphone. Additionally, we divide the data into two sets. The first set has the recorded stereo file as 1 channel, and the other having the audio channels separated for the top and bottom microphone, which will be detailed in section \textbf{5}.

\subsubsection{Training Dataset}
 The total subjects included in the data collection phases were 4 subjects to have a total of 1920 training audio files (.wav) and spectrogram images of such audio files. This dataset was used in the training and validation of the CNN models.

 \subsubsection{Testing Dataset}
  The testing dataset includes 576 audio files and spectrogram images, and 3 subjects collected it (other than those in the training dataset). The dataset size is 576 because it comprises 30\% of the training dataset. This dataset was used for testing the CNN models.

\section{DEEP LEARNING MODELS}

In this section, we discuss the different CNN models and approaches we used. First, we created our model, which was influenced by the literature \cite{hindawi,ultragesture}, And then, we experimented with transfer learning by using different pre-trained Deep learning models and, we achieved the best accuracy with the Xception CNN model \cite{xception}.

\paragraph{Our Model}
Our model is comprised of 6 layers. The input layer is where the image is inserted as greyscale since the greyscale contained all the needed information without the need for an RGB image format. The input shape is an image of dimensions 100 x 100 x 1. Then, the hidden layers include 5 convolutional layers, each followed by a max-pooling layer. We chose max-pooling since it intensifies the bright pixels in the image where we are interested in the spectrogram image. Finally, we use the Softmax activation function to classify the gestures into probabilities, and the output is the gesture with the highest value. The architecture of the model is shown in Figure \ref{cnn_org}.

\paragraph{Xception Models}
We use the Xception Model created by Google \cite{xception}. Xception performed significantly better compared to our Original CNN Model. We fine-tuned the Xception Model to work with our dataset, and we used the pre-trained ImageNet dataset since it produced better results than the untrained Xception model.

\begin{figure}[htbp]
\centering
\includegraphics[width=7.5cm,height=7.5cm,keepaspectratio]{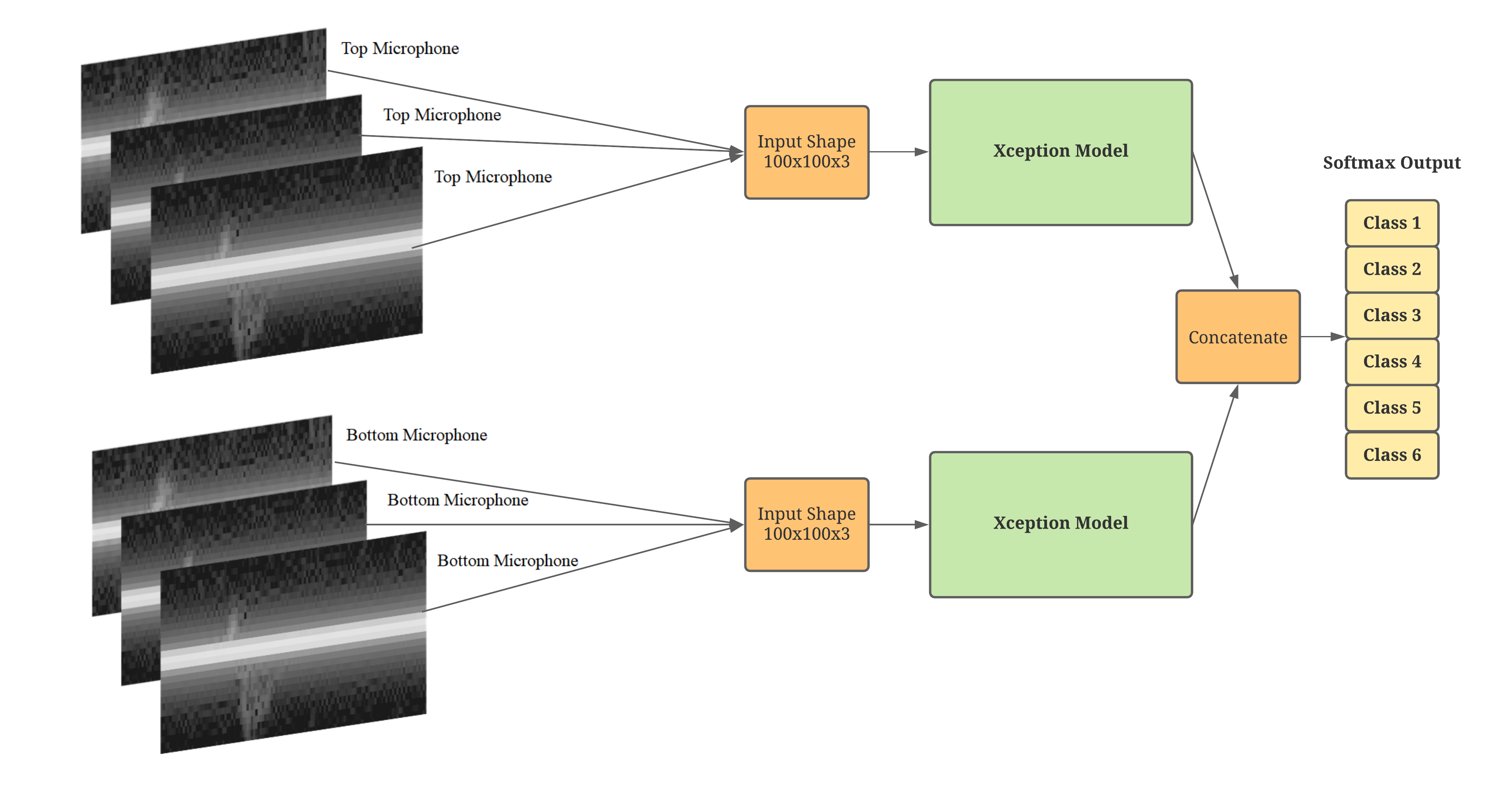}
\caption{Xception Late Fusion architecture}
\label{late_fusion}
\end{figure}

There are three approaches for the Xception Model:
\begin{enumerate}
  \item The first is the Basic CNN Xception Model. Our first attempt was to input the spectrogram as an RGB image due to the nature of the pre-trained model since the Xception model required the input to be of RGB format. In another attempt, we also inputted the spectrogram image as greyscale. However, as three channels were required considering the model takes RGB format, we concatenated the same channel three times to have a grey-scale be a three-channel image to work with the pre-trained model while keeping the greyscale color range. The results showed that concatenating the grey-scale spectrogram image produced better results than the RGB spectrogram image, and thus we opted with the grey-scale image in all our models. 
 
  \item The second is the Early Fusion Xception model. The structure of this model is similar to the Basic CNN Xception Model, and the difference is in the input. The input in this model comes from two sources, the bottom and top microphones. We separated the audio file into two channels (top and bottom), and we converted each audio file to its corresponding spectrogram, as shown in Figure \ref{sep_audio} Hence there are 1920 spectrograms for each channel. Then we concatenated the top microphone, the bottom microphone, and the default spectrogram input, which is the spectrogram converted from the stereo .wav without the separation. Therefore, forming a three-channel image that we can input into the Xception model. Figure \ref{early_fusion}shows the design of this model.
  
\begin{figure}[htbp]
\centering
\includegraphics[width=7cm,height=7cm,keepaspectratio]{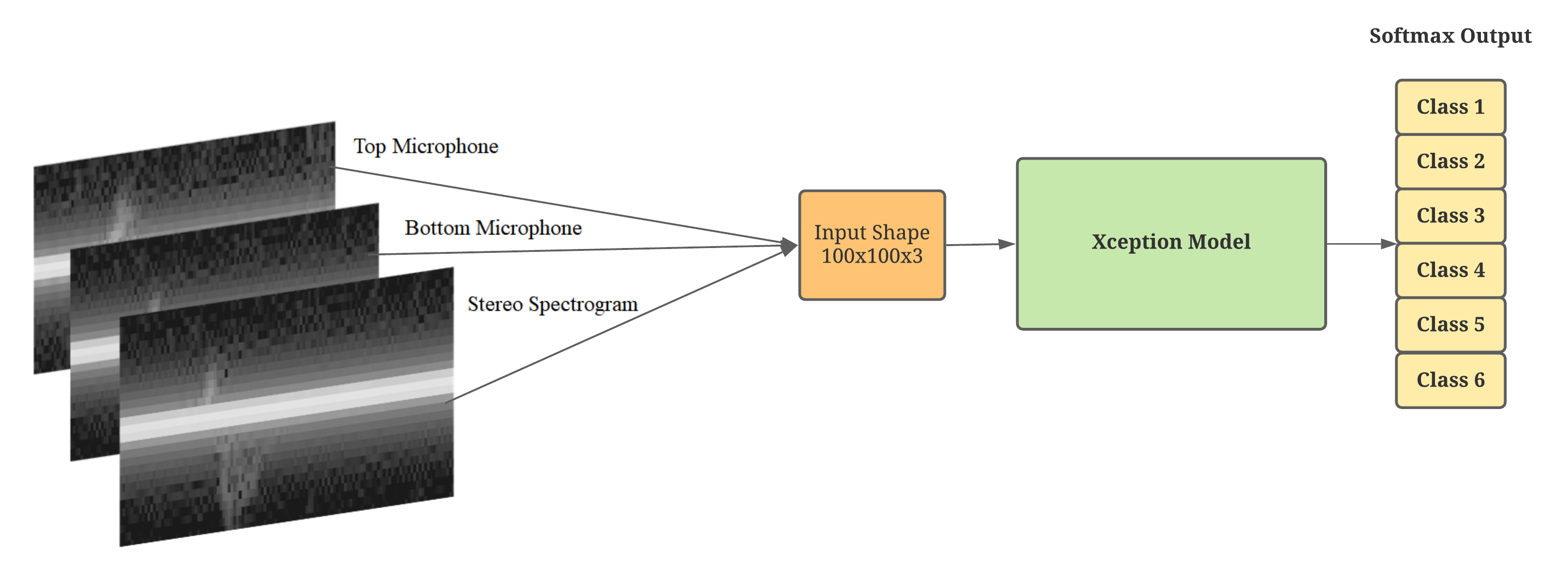}
\caption{Xception Early Fusion architecture}
\label{early_fusion}
\end{figure}
  
  \item The third is the Late Fusion CNN model. In this model, we take the individual spectrogram of the corresponding top and bottom microphones and train each separately. We then concatenate the output of the two models to get the result. For consistency, we used two Xception models, each with the input of a three-channel grey-scale, similar to the first Xception model mentioned above.
  Figure \ref{late_fusion}shows the design of this model.
\end{enumerate}

\section{RESULTS AND DISCUSSION}
\raggedbottom
This section shows the results with different CNN models by showing the achieved accuracy based on the testing dataset, which comprises 30\% of the original training dataset.
\begin{table}[H]
\caption{\textbf{Accuracy (\%) of the different models based on our testing dataset}}

\centering
\begin{tabular}{|l|c|}
\hline
\multicolumn{2}{|c|}{CNN Models Results}                 \\ \hline
\multicolumn{1}{|c|}{Model Name} & Accuracy (\%)        \\ \hline
Original CNN                        & 81.70                 \\ \hline
Xception Model                     & 87.15                 \\ \hline
Xception Late Fusion             & 92.19                \\ \hline
Xception Early Fusion            & \textbf{93.58}        \\ \hline
\end{tabular}
\end{table}

\begin{figure}[htbp]
\centering
\includegraphics[width=4cm,height=4cm,keepaspectratio]{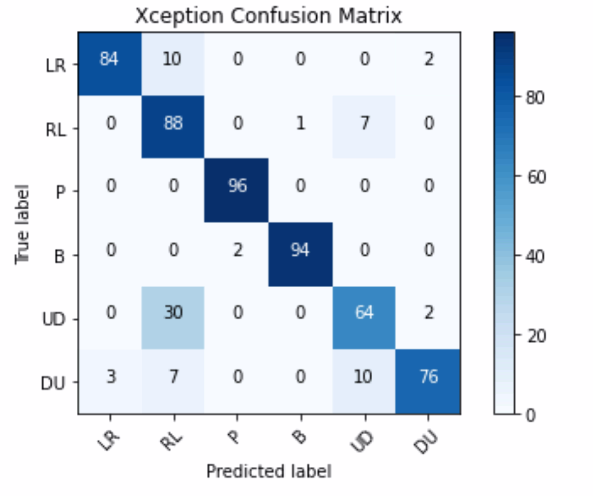}
\caption{Basic Xception Confusion Matrix based on testing dataset. LR: Swipe Right, RL: Swipe Left, P: Push Inwards, B: Block Microphone, UD: Swipe Down, DU: Swipe Up}
\label{confusion_basic}
\end{figure}

\begin{figure}[htbp]
\centering
\includegraphics[width=4cm,height=4cm,keepaspectratio]{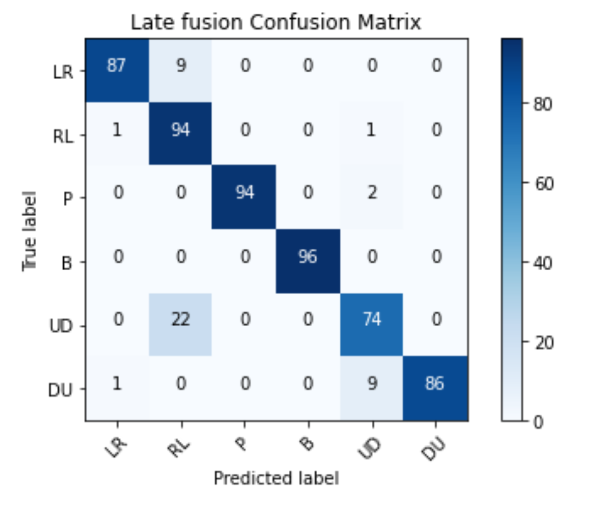}
\caption{Late Fusion Xception Confusion Matrix based on testing dataset}
\label{confusion_late}
\end{figure}

\begin{figure}[htbp]
\centering
\includegraphics[width=4cm,height=4cm,keepaspectratio]{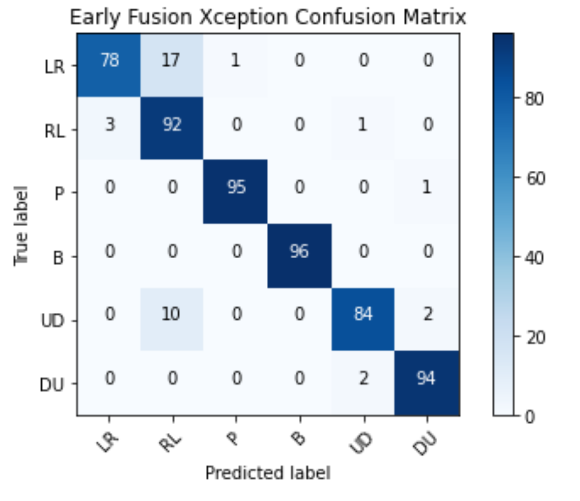}
\caption{Early Fusion Xception Confusion Matrix based on testing dataset}
\label{confusion_early}
\end{figure}

The results showed that using separating the spectrogram for the top and bottom microphone has achieved higher results than using the combined spectrogram for both the top and bottom. This can be shown through the confusion matrices in Figures \ref{confusion_basic}, \ref{confusion_late}, and \ref{confusion_early}, as using top and bottom microphones separately gave a better differentiation when classifying the gestures.

\section{\uppercase{ CONCLUSION}}
\raggedbottom
\label{sec:conclusion}

This paper discusses the techniques used to successfully classify hand gestures using inaudible frequencies emitted from a smartphone. Hence, the methods proposed provide an analysis of the changes that happen in the frequency over time to classify hand gestures. Through testing and comparing ,the three dual-channel input fusion methods provided the best results compared to the single-channel input method. The Early Fusion CNN Model produced the best results compared to the others. It successfully classified 6 hand gestures with an accuracy of 93.58\%.

Moreover, separating the audio files into two separate channels has shown better results and gives more space for more gestures where we will leave to our future work.

\section*{DATA AVAILABILITY}

Our dataset is publicly available via this link:  \url{https://drive.google.com/drive/folders/19Me8bCMPyCm1NOr6rk7apOiCOQXLmdqY}. The training and testing data includes 1,920 and 576 audio files and their corresponding spectrogram images respectively.
\raggedright

\section*{\uppercase{Acknowledgements}}

We would like to thank our supervisors for their immense support and help with our project. Also, we want to thank our volunteers, outside of our team, for helping in the data collection phase: George Maged, Andrew Fahmy, Farida El-Refai, and Ali Madany.

\bibliographystyle{apalike}
{\small
\bibliography{Paper.bbl}}

\end{document}